\documentclass[twocolumn,english,nofootinbib,bibnotes,superscriptaddress,prb]{revtex4-1}

\usepackage[version=3]{mhchem}

\usepackage{graphicx}
\usepackage[usenames,dvipsnames]{color}
\usepackage{dcolumn}
\usepackage{bbm}
\usepackage{bm}
\usepackage{amsmath}
\usepackage{amssymb}
\usepackage{times}
\usepackage{placeins}
\usepackage{hyperref}
\hypersetup{colorlinks=false}

\usepackage{csquotes}

\newcommand{\dif}[1]{\ensuremath{\medspace \mbox{d} #1}}

\newcommand{\fdif}[2]{\ensuremath{\frac{\delta #1}{\delta #2}}}

\newcommand{\p}[1]{\ensuremath{\mathbf{#1}}}

\newcommand{\m}[1]{\ensuremath{\mathrm{#1}}}

\graphicspath{{figures/draft/}}

\begin{document}

\title{Functional renormalization group approach to SU(\textit{N}) Heisenberg models:\\
	Momentum-space RG for the large-\textit{N} limit}

\author{Dietrich Roscher}
\affiliation{Institute for Theoretical Physics, University of Cologne, 50937 Cologne, Germany}
\affiliation{Department of Physics, Simon Fraser University, Burnaby, British Columbia, Canada V5A 1S6}
\author{Finn Lasse Buessen}
\affiliation{Institute for Theoretical Physics, University of Cologne, 50937 Cologne, Germany}
\author{Michael M. Scherer}
\affiliation{Institute for Theoretical Physics, University of Cologne, 50937 Cologne, Germany}
\author{Simon Trebst}
\affiliation{Institute for Theoretical Physics, University of Cologne, 50937 Cologne, Germany}
\author{Sebastian Diehl}
\affiliation{Institute for Theoretical Physics, University of Cologne, 50937 Cologne, Germany}

\date{\today}

\begin{abstract}
In frustrated magnetism, making a stringent connection between microscopic spin models and macroscopic properties of spin liquids remains an important challenge. A recent step towards this goal has been the development of the pseudofermion functional renormalization group approach (pf-FRG) which, building on a fermionic parton construction, enables the numerical detection of the onset of spin liquid states as temperature is lowered. In this work, focusing on the SU($N$) Heisenberg model at large $N$, we extend this approach in a way that allows us to directly enter the low-temperature spin liquid phase, and to probe its character. Our approach proceeds in momentum space, making it possible to keep the truncation minimalistic, while also avoiding the bias introduced by an explicit decoupling of the fermionic parton interactions into a given channel. We benchmark our findings against exact mean-field results in the large-$N$ limit, and show that even without prior knowledge the pf-FRG approach identifies the correct mean-field decoupling channel. On a technical level, we introduce an alternative finite temperature regularization scheme that is necessitated to access the spin liquid ordered phase. In a companion paper [\onlinecite{RSP17}] we present a different set of modifications of the pf-FRG scheme that allow us to study SU($N$) Heisenberg models (using a real-space RG approach) for arbitrary values of $N$, albeit only up to the phase transition towards spin liquid physics.
\end{abstract}


\maketitle

\section{Introduction}
\label{Intro}

In the field of quantum magnetism, the Heisenberg model has long served as one of the most elementary microscopic frameworks to capture a wide range of magnetic phenomena. Of particular interest are Heisenberg antiferromagnets
that easily experience frustration effects by competing exchange interactions or geometric constraints which inhibit
the formation of a conventional N\'eel ordered ground state. Instead, quantum spin liquids \cite{Anderson1973,Lee2008} (QSLs) can form --  collective quantum states, in which the elementary magnetic moments do not order and and remain strongly fluctuating, albeit in a correlated manner \cite{Balents2010,Lhuillier2011,Misguich2011}. 
QSLs have caught the imagination of condensed matter physicists ever since their theoretical inception \cite{Anderson1973} some four decades ago, though it has long remained hard to unambiguously identify these elusive states in candidate materials \cite{Lee2008} or microscopic model systems. 
Today, a quantum spin liquid state is best characterized and positively discriminated from more conventionally ordered ground states by (i) the formation of macroscopic entanglement \cite{LevinWen2006,KitaevPreskill2006} and (ii) the simultaneous emergence of fractionalized degrees of freedom and gauge fields \cite{Savary2017}. 

The connection between a particular spin liquid state and its parent spin Hamiltonian is still fragile, and it remains an open challenge to derive the above characteristic macroscopic phenomena related to spin liquid physics directly from microscopic Heisenberg spin models.
To overcome this obstacle, one would like to explore the properties of candidate spin systems starting from their microscopic description over all length scales down to the very long-wavelength physics, where spin liquid behavior is presumed to occur. 
A natural framework for such a description is provided by renormalization group (RG) methods, where the low-energy phenomenology is systematically approached  upon successively integrating out short-wavelength fluctuations by lowering the RG momentum scale $\Lambda$. 
Specifically for spin models, a pseudofermion functional renormalization group~\cite{Reuther2010}~(pf-FRG) has been developed, 
that applies the established functional renormalization group \cite{Wetterich1993} approach to fermion systems \cite{SalmHon2001,RevModPhys.84.299} on the level of auxiliary (pseudo) fermions introduced in a spin decomposition (or parton construction).
This purely fermionic RG approach treats all interaction channels on equal footing, and therefore avoids a bias which is introduced, for instance, by choosing a specific mean-field decoupling.
The applications of this pf-FRG approach have provided indications for spin-liquid states in a number of frustrated quantum magnets 
\cite{Reuther2010,Reuther2011,Reuther2011a,Reuther2011b,Reuther2011c,Reuther2012,Reuther2014,Reuther2014a,Rousochatzakis2015,Buessen2016,Iqbal2017,Buessen2017}. However, the evidence for spin liquid physics in the pf-FRG approach is typically an indirect one -- such as the absence of a singularity in the RG flow of the spin susceptibility (which would otherwise indicate the onset of magnetic ordering). An unambiguous, positive identification of spin liquid physics as well as a detailed characterization of a putative spin liquid ground state has so far remained elusive within the pf-FRG approach.
The reason for this is twofold: 
 
(1)~A number of spin liquids are not characterized by an order parameter that is {\em bilinear} in the fermion fields, and as such do not exhibit an obvious ordering transition.
Instead their nature reveals itself in fermionic four-point correlation functions and their momentum structures, which however remain subtle and hard to detect. 
Examples of such spin liquids are gapless quantum spin liquids with spinon Fermi surfaces \cite{Motrunich2005,Lee2005,Sheng2009,Block2011}, Majorana Fermi surfaces \cite{Hermanns2014,OBrien2016}, or Bose metals \cite{Sheng2008}.
 
(2)~Bilinearly ordered spin liquids, in contrast, are bound to exhibit a divergence in the fermion four-point function at some finite RG scale $\Lambda_{\m{div}}$ as discussed in a companion paper~\cite{RSP17} of this manuscript. While the divergence itself indicates the onset of a new phase, it inhibits an analysis of the RG flow for scales $\Lambda < \Lambda_{\m{div}}$ and as such a more detailed characterization of the emergent spin liquid. 

A common way to circumvent the second problem is to introduce an order parameter field by means of a Hubbard-Stratonovich transformation (bosonization) and include its correlation functions into the RG analysis. While this approach works very well if the structure of the order parameter is known~\cite{PhysRevB.78.174528}, it introduces a questionable bias if this is not the case. Since spin liquid order parameters can be rather unconventional and often vary only from one another by subtle changes in their symmetry structure, it is much more desirable to have a computational framework that distills this subtle information in the RG flow than vice versa (probing potential order parameters). As such the plain Hubbard-Stratonovich strategy cannot be faithfully applied in the context of spin liquid physics. 

In this manuscript, we will proceed to describe precisely such an unbiased computational approach and implement a pf-FRG scheme that allows to positively identify and characterize spin liquid physics in SU($N$) Heisenberg models. To do so, we adapt an approach first applied in the context of fermionic FRG calculations~\cite{Salmhofer2004} and introduce an infinitesimal symmetry breaking initial condition. The explicit symmetry breaking regularizes the divergence of the fermion correlation function if it addresses the correct symmetry pattern and thus allows to compute infrared observables for scales $\Lambda < \Lambda_{\m{div}}$. Conversely, if the chosen infinitesimal symmetry breaking term was not correct, it will remain infinitesimal throughout the entire flow. As such, our novel approach allows us to unite the advantages of both, the fermionic and bosonized strategies outlined above.

\subsection*{Overview of results}

The main result of the manuscript at hand is an extension of the  pseudofermion functional renormalization group~\cite{Reuther2010} (pf-FRG) formalism to SU($N$) Heisenberg models, for which we demonstrate that, for the first time, we can run the RG flow deep into spin liquid phases.
Example results for the spin-1/2 SU($N$) Heisenberg antiferromagnet on the square lattice substantiate the applicability of this
extended pf-FRG framework and make contact to exact results in the $N\rightarrow\infty$ limit. 

So far, it has  remained elusive to directly access the spin liquid phase within the pf-FRG framework, which in its original real-space formulation (adapted to SU($N$) Heisenberg models in a companion manuscript \cite{RSP17}) is valid only up to the onset of ordering (either in the magnetic or spin liquid sector) where the flow of the RG equations is rendered invalid by a divergence of the fermionic four-point function. 
Here, we devise a simplified truncation scheme that analytically reproduces this behavior in terms of a single running coupling in {\em momentum space}.
To expand the validity of the pf-FRG description deep into the spin liquid phase, we introduce a small explicit symmetry breaking term in the ordering channel that allows to effectively regularize the divergence. 
Essential information about zero temperature observables, such as the gap and magnetic susceptibility, is then obtained from the solution of only two RG equations. Furthermore, we exhibit the absence of long-range magnetic ordering by showing that magnetic order parameters neither grow nor regularize the divergence.
The bias introduced by this procedure is thus under systematic control. 
As opposed to the Hubbard-Stratonovich approach, no fluctuating boson field is introduced, and the description of the system is entirely in terms of fermionic fields and their fluctuations. We show explicitly that this implies a positive discrimination between possible ordering channels, since an inappropriate choice of the symmetry breaking pattern does not regularize the fermionic divergence and thus predicts its own unsuitability. 

In previous pf-FRG approaches, a cutoff scheme has been proposed, where temperature and running scale parameter could be identified with each other via a simple linear relation. 
Here we show that this relation breaks down inside the ordered phase.
Instead we propose to access the finite temperature regime by an alternative cutoff scheme. 
By this procedure, we are able to exactly reproduce mean-field observables for any given temperature. The application of this cutoff scheme furthermore opens up a strategy to exactly implement fermion number constraints by means of Popov-Fedotov chemical potentials.


The discussion of our results in the remainder of this manuscript is structured as follows. 
In Sec.~\ref{AAMF}, we introduce the staggered flux spin liquid, a well-understood and simple model system that will be used to benchmark our pf-FRG extension. 
Sec.~\ref{SymmFRG} introduces a pf-FRG approach in  momentum-space  tailored for simplicity and quick analysis of symmetry breaking patterns. It is then benchmarked in the symmetric regime. In Sec.~\ref{SSBFRGT0}, the symmetry broken regime is accessed within a zero-temperature formalism, which already partially reproduces the results from Sec.~\ref{AAMF}.
Furthermore, the unbiasedness of the method is demonstrated by analyzing a different symmetry breaking pattern. 
As a limitation of this $T=0$ approach, we exhibit that the linear relation between the RG scale parameter and physical temperature breaks down inside of the spin liquid phase. Therefore, finite temperature observables cannot be extracted reliably. 
To overcome this limitation, Sec.~\ref{SSBFRGTf} then addresses the problem of finite-temperature observables inside the symmetry broken regime by implementing a finite-temperature cutoff scheme.
Finally, we demonstrate how the fermion number constraint, that provides the relation to the original spin model, is implemented exactly within our pf-FRG scheme by means of the Popov-Fedotov method. 
Conclusions are finally drawn in Sec.~\ref{Conclusions}, followed by some appendices providing details of the computations.

\section{Model system: Staggered-flux spin liquid}
\label{AAMF}

To establish controlled access to spin liquid phases by means of pf-FRG, we study a well-understood realization of the Heisenberg model. Here, we introduce our model and briefly recapitulate known results~\cite{AffleckMarston88,ArovasAuerbach88}.
Our starting point is the Hamilton operator of the nearest-neighbor SU($N$)-symmetric Heisenberg model on the two-dimensional square lattice
\begin{equation}
\label{HeisenHam}
\mathcal{H} = \frac{J}{N}\sum_{\langle ij \rangle} \mathbf{S}_i\cdot\mathbf{S}_j
\end{equation}
where the components $S^\mu_i$ of the SU($N$) spin operators $\mathbf{S}_i$ carry the site indices $i,j$ and $\mu \in \{1,...,N^2-1\}$. The spins interact via the antiferromagnetic exchange coupling $J>0$.

For a concrete calculation, a symmetry group and representation for the spin operators $S^\mu_i$ must be chosen. The generic SU(2) group in its fundamental representation is known to exhibit a (classical) antiferromagnetically ordered ground state at vanishing temperature~\cite{ManousakisRev91}. Several different symmetry groups and representations have been investigated in the past~\cite{Read89,Read1991,BaezLargeS17} in order to suppress and enhance different types of fluctuations, eventually leading to different ground states as well. For SU$(N)$ spins in their fundamental spin-$\frac{1}{2}$ representation, it was found that the classical antiferromagnet can give way to a \enquote{staggered flux spin liquid} phase~\cite{AffleckMarston88} for $N\rightarrow\infty$. 
In fact, this result is intimately tied to a representation of the spin operators by Abrikosov pseudofermions~\cite{Abrikosov1965}
\begin{equation}
\label{PFDef}
S^\mu_i = f^\dagger_{i\alpha} T^\mu_{\alpha\beta} f^{\phantom\dagger}_{i\beta}, \quad f^\dagger_{i\alpha}f^{\phantom\dagger}_{i\alpha} = \frac{N}{2},
\end{equation}
with $T^\mu$ being the $N$-dimensional representation of the generators. 
Summation over equal spin indices is implied. 
The number constraint in Eq.~\eqref{PFDef} ensures that the Hilbert space of the pseudofermions is not enlarged w.r.t the one of the original spin operators.

The partition function of the model can now be written as $\mathcal{Z}=\int\mathcal{D}[f,\lambda]\exp\{-S[f]+S_c[\lambda]\}$ with the action
\begin{align}\label{eq:pfaction}
	S[f]=& \int_\tau\Big[\sum_i f^\dagger_{i\alpha}\dot{f}_{i\alpha}-\frac{J}{N}\sum_{\langle ij \rangle}f^\dagger_{i\alpha}f^{\phantom\dagger}_{j\alpha}f^\dagger_{j\beta}f^{\phantom\dagger}_{i\beta}\Big]\,,
\end{align}
where $\int_\tau:=\int_0^\beta \dif \tau$ and we used commutation relations to rearrange the pseudofermions in the four-fermion interaction coming from the Hamiltonian in Eq.~\eqref{HeisenHam}.
Further, we defined
\begin{equation}
\label{lambdaconstraint}
S_c[\lambda] = i\int_\tau\sum_i\lambda_i\Big(f^\dagger_{i\alpha}f^{\phantom\dagger}_{i\alpha} - \frac{1}{2}\Big)\,,
\end{equation}
to impose the fermion number constraint, cf.~Eq.~\eqref{PFDef}.
The four-fermion interaction term can be decoupled by means of a Hubbard-Stratonovich transformation, introducing the spin-singlet site-nonlocal boson field $Q_{ij}\sim f^\dagger_{j\alpha}f_{i\alpha}$. 
It can be seen as an order parameter field for the \emph{local} U$(1)$ symmetry that was introduced \emph{artificially} when representing the Heisenberg Hamiltonian in terms of the pseudofermions.

\begin{figure}[t]
\centering
\includegraphics[width = 90mm]{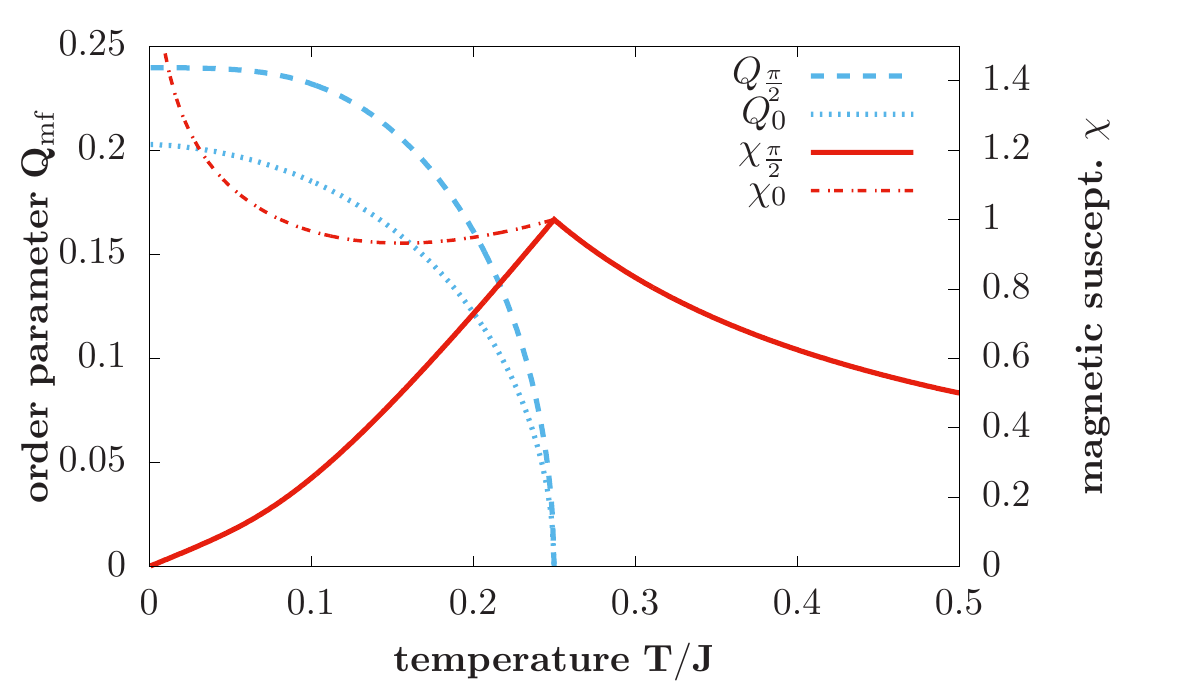}
\caption{{\bf Mean-field order parameter} $Q_{\m{mf}}$ 
and {\bf magnetic susceptibility} $\chi$ 
as functions of temperature. 
For $\varphi=0$ the order parameter is slightly smaller than for $\varphi = \pi/2$ while both solutions share the same critical temperature $T_c = J/4$. 
The zero-temperature magnetic susceptibility vanishes for $\varphi = \pi/2$ whereas it exhibits a logarithmic divergence for $\varphi=0$.}
\label{MFResPlot}
\end{figure}

After Hubbard-Stratonovich transformation, the partition function reads 
\begin{align}\label{MFZ}
\mathcal{Z} &= \int\mathcal{D}[f,Q,\lambda]\exp\Bigg\{\int_\tau  \Bigg[-\sum_i f^\dagger_{i\alpha}\dot{f}_{i\alpha}\\
&
+\sum_{\langle ij \rangle}\Big(\frac{N}{J} Q^\dagger_{ij}Q_{ij} + (Q^\dagger_{ij}f^\dagger_{j\alpha}f^{\phantom\dagger}_{i\alpha} + \text{h.c.})\Big)\Bigg] + S_c[\lambda]\Bigg\}.\nonumber
\end{align}
Since the action is now quadratic in the fermion fields, the latter can be integrated out, resulting in a new action $S_Q$ and the gap equations 
\begin{equation}
\fdif{S_Q}{Q^\dagger} = \fdif{S_Q}{Q} = \fdif{S_Q}{\lambda} = 0.
\end{equation}
While the solution for the constraint field $\lambda = 0$ can be obtained straightforwardly, $Q$ has multiple valid solutions for the 2D square lattice. It turns out that 
\begin{equation}
	Q^\dagger_{ij} = Q e^{i\theta},\, Q_{ij} = Q e^{-i\theta}\ \ \rm{with}\ \  
	\theta = \left\{\begin{matrix}   \varphi, & \p{e}_{ij} \sim \p{e}_x \\ 0, & \p{e}_{ij}\sim \p{e}_{y} \end{matrix} \right.
\end{equation}
gives a lower free energy for $\varphi = \frac{\pi}{2}$ (flux phase, spin liquid) than for $\varphi = 0$ (uniform \enquote{BZA} phase)~\cite{BZA1987} in the symmetry broken regime. The mean-field values for $Q$ and the magnetic susceptibility, which is defined as
\begin{equation}
\chi = \int_0^\beta\dif{\tau}\dif{\tau'}\langle S^z_{i,\tau} S^z_{j,\tau'} \rangle\,,
\end{equation}
are shown as functions of temperature in Fig.~\ref{MFResPlot}.

A couple of peculiarities should be noted about these results, which will be important for our further discussion. Firstly, in the limit $N\rightarrow\infty$, the mean-field gap equation becomes exact. This is due to the fact that higher-order fluctuations are suppressed\cite{AuerbachBook} by powers of $N$. As a consequence, the Mermin-Wagner-Hohenberg theorem~\cite{MerminWagner,Hohenberg1967} prohibiting true long-range order in two spatial dimensions at finite temperature is bypassed~\cite{ThiesSchoen}.
Secondly, the magnetic susceptibility does not diverge at the critical temperature where $Q > 0$. This indicates that it is, indeed, not the global SU$(N)$ symmetry which is spontaneously broken but rather the local U$(1)$ symmetry\footnote{Here, we follow previous literature in employing the term ``symmetry breaking'' in the context of local gauge symmetries and use it analoguous as, e.g., for mean field superconductors: There is no spontaneous symmetry breaking witnessed by correlators that are not gauge invariant, as a consequence of Elitzur's theorem. However, there are observable consequences due to the onset of ordering witnessed by gauge-invariant quantities, such as the expression of a gap in a superconducting phase.}.
Thirdly, while the uniform solution with $\varphi = 0$ is energetically not preferred, it shares the same critical temperature with the actual ground state at $\varphi=\frac{\pi}{2}$. Thus, the physics of local U$(1)$ symmetry breaking may already be captured by the simplest order parameter structure, whereas the additional breaking of translation invariance and the occurrence of actual spin liquid physics is tied to the inherently nonlocal realization of $Q$. Finally, the fermion number constraint is, at this level of approximation, effectively lost since $\lambda = 0$. 
Technically, it is therefore not clear whether the results, e.g., in Fig.~\ref{MFResPlot} are in fact solutions of the original spin problem from Eq.~\eqref{HeisenHam} or rather an unconstrained fermion system and whether this makes a difference physically. We will revisit this question in Sec.~\ref{Popotov}, below.

\section{Momentum-space FRG: symmetric regime}
\label{SymmFRG}

Let us now set up a pf-FRG scheme which fully incorporates the exact large-$N$ results from Sec.~\ref{AAMF}. Part of this scheme has already been realized in Ref.~\onlinecite{RSP17}, where the symmetric regime was investigated. The goal of this work is to extend the pf-FRG into the symmetry broken regime by introducing symmetry breaking initial conditions.
It is, of course, also possible to implement the latter in the position space pf-FRG scheme used in Ref.~\onlinecite{RSP17}. 
Here, we do not follow this path mainly for two reasons: 
Firstly, the numerical cost for even the simplest possible realization of a spatially non-local order parameter as required by the spin liquid description in Sec.~\ref{AAMF} is very high.
Secondly, in more realistic cases, it is in general not obvious what the particular structure of the prospective spin liquid order parameter is. One would therefore have to test different order parameter structures to find the physically realized one, further increasing the cost. 

We therefore strive to develop a simplified framework which provides the opportunity to conduct comparatively quick and cheap tests. Furthermore, such a framework provides additional analytical insight as the reduction to only a few running couplings is bound to expose the decisive physical ingredients.
To that end, we apply momentum space functional renormalization group for the flowing effective average action $\Gamma_\Lambda$ as given by the Wetterich equation~\cite{Wetterich1993,SalmHon2001}
\begin{equation}
\label{Wetterhofer}
\partial_\Lambda \Gamma_\Lambda = \frac{1}{2}\int_\tau\sum_if^\dagger\partial_\Lambda\mathcal{P}_\Lambda f + \frac{1}{2}\m{STr}\left[\frac{\partial_\Lambda \mathcal{P}_\Lambda}{\Gamma_\Lambda^{(2)}}\right].
\end{equation}
Here,  $\Lambda$ is the RG scale parameter. For $\Lambda\rightarrow\Lambda_{\m{UV}}$, it is identical to the microscopic action given in eq.~\eqref{MFZ}. For $\Lambda\rightarrow 0$, it becomes the full effective action of the model. $\mathcal{P}_\Lambda$ is the free inverse propagator, including a (multiplicative) regularization function (see below).
The FRG equation~\eqref{Wetterhofer} itself is exact. In order to solve it, however, we have to employ approximations. Primarily, this means a restriction of $\Gamma_\Lambda$ with respect to the infinite number of terms which are compatible with the symmetries of the model.

Before we implement symmetry breaking initial conditions, we consider a symmetric version of our simplified framework in order introduce our simplifications and check, whether results from real space pf-FRG such as the critical temperature are reproduced correctly.
Our ansatz for the effective average action follows the action defined in Eq.~\eqref{eq:pfaction} and in frequency-momentum space is given by
\begin{align}
\label{T0SymmGk}
\Gamma_\Lambda^{\m{sym}} &=\int_{1\in \text{FZ}} f^\dagger_{1\alpha} \mathcal{P}_\Lambda f^{\phantom\dagger}_{1\alpha} - \frac{J_\Lambda}{N} \int_{1...4\in \text{FZ}}\Big[\cos(k_{2,x} - k_{3,x})\nonumber\\
&\quad + \cos(k_{2,y} - k_{3,y})\Big]f^\dagger_{1\alpha}f^{\phantom\dagger}_{2\alpha}f^\dagger_{3\beta}f^{\phantom\dagger}_{4\beta}\,\delta_{1234}\,,
\end{align}
where $1...4$ correspond to combined frequency and momentum variables $(\omega_1,k_1),...,(\omega_4,k_4)$ and we have introduced the shorthand notations $\int_{1 \in \text{FZ}}:=\int_{\omega_1}\int_{k_1 \in \text{FZ}}$ for the frequency and momentum integration as well as $\delta_{1234}:=\delta(1-2+3-4)$.
The momenta are integrated over the full first Brillouin zone (FZ) of the square lattice and we limit ourselves to $T=0$ for the time being.
Furthermore, the free propagator
\begin{align}
\mathcal{P}_\Lambda = i\,\omega\cdot\vartheta_\Lambda^{-1}
\end{align}
is equipped with the  inverse multiplicative cutoff function~$\vartheta_\Lambda^{-1}$ which will be specified below.
 
Compared to the position-space ansatz of Ref.~\onlinecite{RSP17}, drastic simplifications have been made: the four-point function is reduced to one single running coupling $J_\Lambda$ independent of frequency and momentum. The running of the two-point function is neglected. As a result, the unique projection rule for the $\beta$~function of the four-point function is particularly simple,
\begin{equation}
\label{T0SymmJProj}
\partial_\Lambda J_\Lambda (f^\dagger_\alpha f^{\phantom\dagger}_\alpha f^\dagger_\beta f^{\phantom\dagger}_\beta)\Omega = -\frac{N}{2}\partial_\Lambda \Gamma_\Lambda^{\m{sym},4f}\,.
\end{equation}
Here, $\Omega$ is the space-time volume and $\Gamma_\Lambda^{\m{sym},4f}$ is the four-fermion sector of the effective average action or the Wetterich equation, respectively. $\partial_\Lambda J_\Lambda$ is then obtained by comparison of coefficients. Furthermore, the projection rule is evaluated for $f_{a\alpha}$ at vanishing frequencies and momenta, implemented by $f^{(\dagger)}_{a\alpha} = f^{(\dagger)}_\alpha\delta(\omega_a)\delta(k_a)$. This simplifies the non-trivial momentum structure of the four-fermion sector.

Employing a fluctuation expansion of the Wetterich equation, the large-$N$ $\beta$ function can be extracted as
\begin{equation}
\label{T0SymmBeta}
\partial_\Lambda J_\Lambda = -J_\Lambda^2\partial_\Lambda \int_\omega \frac{\vartheta_v^2}{(i\omega)^2} = -\frac{J_\Lambda^2}{\pi \Lambda^2}.
\end{equation}
In order to arrive at the last identity in eq.~\eqref{T0SymmBeta}, the cutoff function $\vartheta_\Lambda$ had to be specified. Since we work at $T=0$, we employ a zero-temperature step function cutoff
\begin{equation}
\label{StepReg}
\vartheta_\Lambda = \Theta(|\omega|-\Lambda), \quad \partial_\Lambda\vartheta_\Lambda = -\delta(|\omega|-\Lambda).
\end{equation}
Eq.~\eqref{T0SymmBeta} can now be integrated directly, yielding
\begin{equation}
\label{T0SymmSoln}
J_\Lambda = \frac{\pi \Lambda}{\pi \Lambda - 1}.
\end{equation}
If not stated otherwise, the initial value will be set to $J_{\Lambda\rightarrow\Lambda_{\m{UV}}\rightarrow\infty} = 1$ for convenience. The breakdown of the flow at $\Lambda_{\m{sb}} = \pi^{-1}$ indicates the onset of spontaneous symmetry breaking.

Since the cut-off frequency integral can be interpreted as an approximation to a Matsubara sum, it is possible to identify the symmetry breaking scale $\Lambda_{\m{sb}}$ with a critical temperature~\cite{Reuther2011c}. This identification can, at least in the symmetric phase, be used to calculate observables such as the magnetic susceptibility. The precise conversion factor, however, may depend on the structure of the $\beta$-function and should therefore be determined case by case, see below.

The mean-field results, cf. Fig.~\ref{MFResPlot}, we want to compare to are obtained with the fermion number constraint released completely. Thus, we can also perform an FRG calculation at finite fermion Matsubara temperature. In order to do so, we have to choose a different cutoff function, since Eq.~\eqref{StepReg} is ill defined for discrete frequencies. Instead, we can use a cutoff proposed in Ref.~\onlinecite{EnssThesis}, given by
\begin{equation}
\label{EnssReg}
\vartheta_\Lambda = \left\{ \begin{matrix} 0 & |\omega_n| < \Lambda-\pi T \\ \frac{1}{2} + \frac{|\omega_n|-\Lambda}{2\pi T} & \Lambda-\pi T \leq |\omega_n| \leq \Lambda + \pi T\\ 1 & \Lambda + \pi T \leq |\omega_n| \end{matrix}\right..
\end{equation}
The scale derivative of the cutoff, $\partial_\Lambda \vartheta_\Lambda$, selects only two Matsubara frequencies $\omega_n = \pm \pi T \left(2\left\lfloor \frac{\Lambda}{2\pi T}\right\rfloor + 1\right)$ for each particular $\Lambda$ almost everywhere, except for discrete points where it jumps. Here, $\lfloor x \rfloor$ is our notation for the floor operation on $x$. The $\beta$-function is then given by
\begin{equation}
\label{TFSymmSoln}
\partial_\Lambda J_\Lambda = -\frac{J_\Lambda^2}{\pi^3 T^2}\frac{\left(2 + 2\left\lfloor \frac{\Lambda}{2\pi T}\right\rfloor - \frac{\Lambda}{\pi T}\right)}{\left(2\left\lfloor\frac{\Lambda}{2\pi T}\right\rfloor + 1\right)^2}.
\end{equation}
The r.h.s. of this equation is not continuous and we solve it numerically. We find that for $T < \frac{1}{4}$, the flow of $J_\Lambda$ diverges at a finite scale $\Lambda_{\m{div}}$, indicating the onset of spontaneous symmetry breaking. This value for the critical temperature
agrees with the mean-field findings~\cite{ArovasAuerbach88} from Sec.~\ref{AAMF}. We therefore determine the conversion factor between RG scale and temperature in this case to be $\Lambda_{\m{stop}} = \frac{4}{\pi}T$. Here, $\Lambda_{\m{stop}}$ is the scale where the flow has been stopped. This may be due to a divergence during the flow or deliberately for the computation of observables. For the latter case, the correspondence between $\Lambda$ and $T$ is assumed to hold in general, see, e.g., Refs.~\onlinecite{Iqbal2016,RSP17} for a more detailed discussion. We will revisit this interpretation of $\Lambda$ in sec.~\ref{SLC} below.

\section{Symmetry broken regime I: T=0}
\label{SSBFRGT0}

\subsection{Spin-liquid channel}
\label{SLC}

\begin{figure}[t!]
\centering
\includegraphics[width = 0.9\columnwidth]{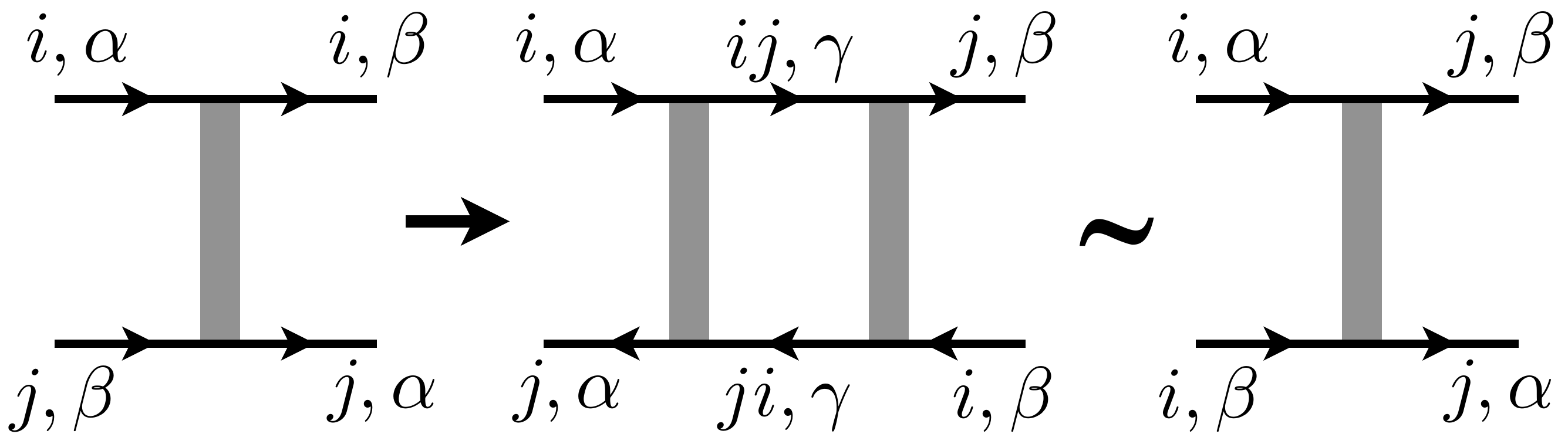}
\caption{{\bf Generation of new vertex structure.} Diagrammatic representation of the process responsible for the generation of the vertex in~Eq.~\eqref{cIVert} out of the original spin interaction $\propto J_\Lambda$ (left vertex). Due to non-locality of the fermion propagator for finite $Q_{\Lambda\rightarrow\infty}$, cf. Eq.~\eqref{GESB}, the vertex in Eq.~\eqref{cIVert} receives finite contributions during the flow, due to loop contributions represented by the diagram in the middle of the figure. Therefore, a finite vertex, Eq.~\eqref{cIVert}, is generated (right vertex), which has to be taken into account in the RG flow.}
\label{cIVertGen}
\end{figure}

Let us now proceed further by exploiting the simplicity of the momentum space approach and venture into the regime of spontaneously broken symmetry. In Ref.~\onlinecite{Salmhofer2004}, an FRG framework was developed for the reduced BCS model which achieves precisely that. We will demonstrate its capabilities directly for the square lattice antiferromagnet at hand.
To that end, we introduce the term
\begin{equation}
\label{GESB}
\Gamma_\Lambda^{\m{esb}} = \int_\tau\sum_{\langle ij \rangle}Q_\Lambda \left[f^\dagger_{i\alpha}f^{\phantom\dagger}_{j\alpha} + f^\dagger_{j\alpha}f^{\phantom\dagger}_{i\alpha}\right]\,,
\end{equation}
into the ansatz for the effective average action $\Gamma^{\m{sb}}_\Lambda = \Gamma_\Lambda^{\m{sym}} + \Gamma_\Lambda^{\m{esb}}$ which breaks the local U$(1)$ symmetry explicitly. Importantly, $\Gamma_\Lambda^{\m{esb}}$ is \emph{not} introduced by means of a Hubbard-Stratonovich transformation. The four-fermion interaction term $\propto J_\Lambda$ is present in $\Gamma_\Lambda^{\m{sb}}$ and unaltered with respect to $\Gamma_\Lambda^{\m{sym}}$. In the limit $Q_{\Lambda\rightarrow\infty} = 0$, the symmetric ansatz from Eq.~\eqref{T0SymmGk} is recovered.

For any finite $Q_{\Lambda\rightarrow\infty}$, the term $\Gamma_\Lambda^{\m{esb}}$ effectively becomes a contribution to the two-point correlation function or self-energy. 
Building on Ref.~\onlinecite{Salmhofer2004}, it was shown in Ref.~\onlinecite{RSP17} that a truncation of $\Gamma_\Lambda$ including terms up to fourth order in the fermion fields \emph{and} involving Katanin's improvement scheme~\cite{KataninSeminar04}, the self-consistent gap equation for the order parameter can be reproduced exactly from the RG flow equations.  Since $Q_\Lambda$ is effectively nonlocal in space, the considerations from Ref.~\onlinecite{RSP17} are directly applicable. We can therefore expect, that $Q_{\Lambda\rightarrow 0} = Q_{\m{mf}}(T=0)$, thus reproducing the mean-field result for the magnitude of the order parameter from Sec.~\ref{AAMF}. Furthermore, the presence of a finite $Q_\Lambda$ is expected to regularize the  finite-$\Lambda$ divergence of the running fermionic coupling(s) due to spontaneous symmetry breaking: since the symmetry is already broken explicitly, no non-analytic behavior can occur.

With respect to the quartic fermion sector of the simplified ansatz in Eq.~\eqref{T0SymmGk}, there is a subtlety to be considered. By introducing the self-energy term, Eq.~\eqref{GESB}, the fermion propagator becomes effectively nonlocal (bilocal) in position space. However, the simplified ansatz $\Gamma_\Lambda^{\m{sym}} + \Gamma_\Lambda^{\m{esb}}$ does not fully account for this so far. 
In particular, the process depicted in Fig.~\ref{cIVertGen} contributes to leading order at large $N$ and generates the new vertex structure
\begin{align}
\Gamma_\Lambda^{I_\Lambda} &= \frac{I_\Lambda}{N}\int_\tau\sum_{\langle ij \rangle} f^\dagger_{i\alpha} f^{\phantom\dagger}_{j\alpha}f^\dagger_{i\beta}f^{\phantom\dagger}_{j\beta} \\
&= \frac{I_\Lambda}{N}\int_{1...4 \in \text{FZ}} f^\dagger_{1\alpha}f^{\phantom\dagger}_{2\alpha}f^\dagger_{3\beta}f^{\phantom\dagger}_{4\beta}\Big[\cos(k_{2,x}+k_{4,x}) \nonumber\\
&\quad+ \cos(k_{2,y}+k_{4,y})\Big]  \delta_{1234}\,.\label{cIVert}
\end{align}
Comparing the momentum space vertices in Eqs.~\eqref{T0SymmGk} and~\eqref{cIVert}, it becomes clear that their only difference is provided by the momentum structure itself. The projection onto specific momentum structures is rather cumbersome and we therefore follow a different path here:
The lattice geometry and symmetry of the Heisenberg interaction, cf. Eq.~\eqref{HeisenHam}, permit a partition of the original square lattice into two sublattices~\cite{ArovasAuerbach88} $A$ and $B$.
Making this partition explicit by introducing spinors $\Psi^\dagger_{1,\alpha} = (f^{\dagger,A}_{1,\alpha},f^{\dagger,B}_{1,\alpha})$, a new $2\times 2$ sublattice space is created, parametrized by the matrices $\mathbbm{1},\sigma_{x/y/z}$ or $\sigma_{\pm} = \frac{1}{2}(\sigma_x\pm i\sigma_y)$, respectively. 
The full momentum space ansatz for the effective average action is then given by
\begin{widetext}
\begin{equation}
\label{T0sbGk}
\begin{aligned}
\Gamma_\Lambda =& \int_{1\in \text{LZ}}\left\{ i\omega_1\left(\Psi^\dagger_{1\alpha}\mathbbm{1}\Psi^{\phantom\dagger}_{1\alpha}\right) + 2Q_\Lambda\left(\Psi^\dagger_{1\alpha}\sigma_x\Psi^{\phantom\dagger}_{1\alpha}\right)\left[\cos(k_x) + \cos(k_y)\right] \right\}\\
&- 2\frac{J_\Lambda}{N}\int_{1...4\in \text{LZ}}\left(\Psi^\dagger_{1\alpha}\sigma_+\Psi^{\phantom\dagger}_{2\alpha}\right)\left(\Psi^\dagger_{3\beta}\sigma_-\Psi^{\phantom\dagger}_{4\beta}\right) \left[\cos(k_{2,x}-k_{3,x}) + \cos(k_{2,y}-k_{3,y})\right]\delta_{1234}\\
&- \frac{I_\Lambda}{N}\int_{1...4\in \text{LZ}}\left(\Psi^\dagger_{1\alpha}\sigma_+\Psi^{\phantom\dagger}_{2\alpha}\right)\left(\Psi^\dagger_{3\alpha}\sigma_+\Psi^{\phantom\dagger}_{4\alpha}\right) \left[\cos(k_{2,x}+k_{4,x}) + \cos(k_{2,y}+k_{4,y}) \right]\delta_{1234}\\
&-\frac{I_\Lambda}{N}\int_{1...4\in \text{LZ}}\left(\Psi^\dagger_{1\alpha}\sigma_-\Psi^{\phantom\dagger}_{2\alpha}\right)\left(\Psi^\dagger_{3\alpha}\sigma_-\Psi^{\phantom\dagger}_{4\alpha}\right) \left[\cos(k_{1,x}+k_{3,x}) + \cos(k_{1,y}+k_{3,y}) \right]\delta_{1234}\,,
\end{aligned}
\end{equation}
\end{widetext}
where LZ in Eq.~\eqref{T0sbGk} denotes the first Brillouin zone of either one of these two sublattices (\enquote{little zone}).

By means of the sublattice ansatz, the two vertices $\sim J_k$ and $\sim I_k$ now become algebraically distinguishable due to their different structure in sublattice space. The projection rules for the running couplings are
\begin{align}\label{T0sbProj}
(\partial_\Lambda Q_\Lambda) (\Psi^\dagger_\alpha \sigma_x\Psi^{\phantom\dagger}_\alpha) &= \frac{1}{4\Omega}\partial_\Lambda \Gamma_\Lambda^{\sigma_x},\\
(\partial_\Lambda J_\Lambda) (\Psi^\dagger_\alpha\sigma_+\Psi^{\phantom\dagger}_\alpha)(\Psi^\dagger_\beta\sigma_-\Psi^{\phantom\dagger}_\beta) &= -\frac{N}{4\Omega}\partial_\Lambda \Gamma_\Lambda^{\sigma_+\sigma_-}, \nonumber%
\\
(\partial_\Lambda I_\Lambda) 
\left[(\Psi^\dagger_\alpha\sigma_+\Psi^{\phantom\dagger}_\alpha)^2+(\Psi^\dagger_\alpha\sigma_-\Psi^{\phantom\dagger}_\alpha)^2\right] &= -\frac{N}{2\Omega}\partial_\Lambda \Gamma_\Lambda^{\sigma_\pm\sigma_\pm}. \nonumber%
\end{align}
The actual flow equations can, again, be extracted by comparison of coefficients.

\begin{figure}[t!]
\centering
\includegraphics[width = 90mm]{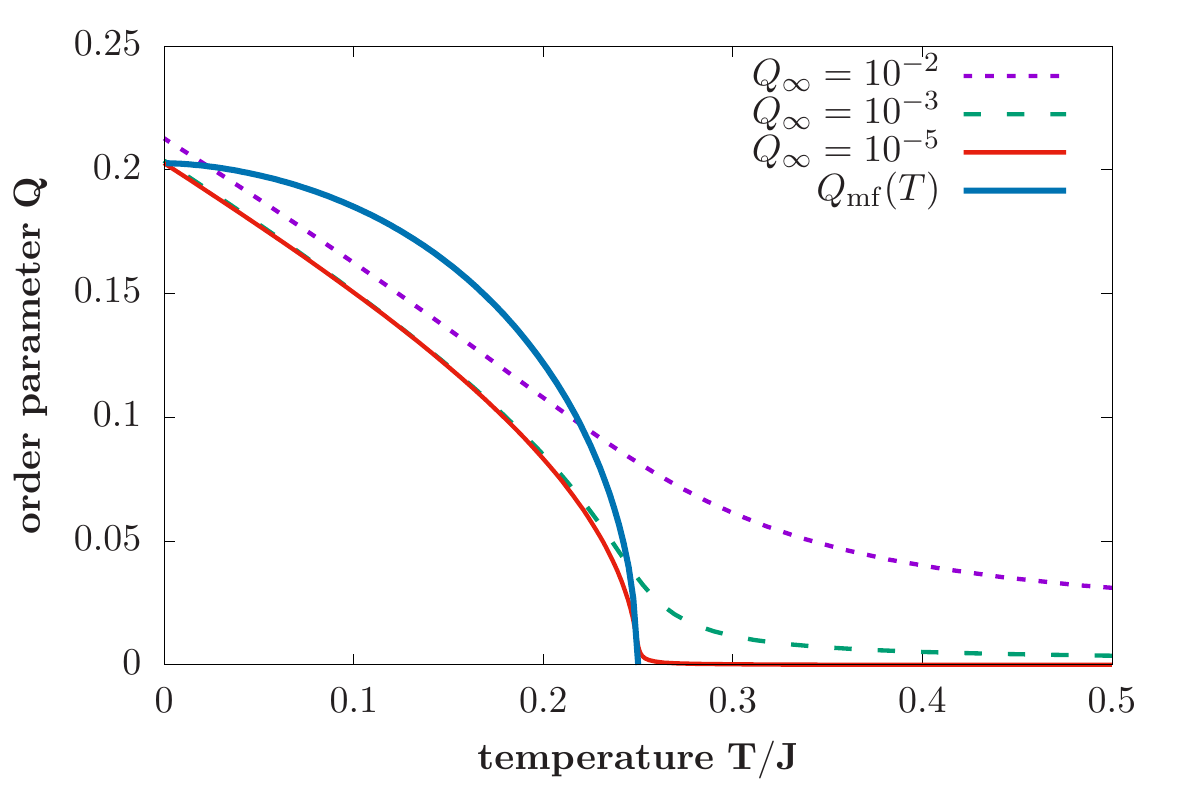}
\caption{Comparison of the {\bf pf-FRG order parameter} magnitude for different initial values $Q_{\Lambda\rightarrow\infty}$ to the mean-field result $Q_{\m{mf}}$, assuming the conversion between temperature and RG scale, $\Lambda_{\m{stop}} = \frac{4}{\pi}T$.}
\label{RGGap}
\end{figure}

A straightforward calculation, cf. App.~\ref{FlowForce}, now leads to the respective $\beta$ functions. More conveniently, considering
\begin{equation}
\label{CoupTrafo}
X_\Lambda = J_\Lambda + I_\Lambda,\quad Y_\Lambda = J_\Lambda - I_\Lambda,
\end{equation}
instead of the original couplings, the flow equations for $X_\Lambda$ and $Y_\Lambda$ decouple. This greatly simplifies further analyses, in particular, as the flow of $Q_\Lambda$ is independent of $Y_\Lambda$ as well. The $\beta$~functions are finally given by
\begin{align}
\partial_\Lambda Q_\Lambda &= -\frac{X_\Lambda}{\pi}\int_{p\in \text{FZ}}\frac{Q_\Lambda\alpha^2}{4Q_\Lambda^2\alpha^2+\Lambda^2},\label{QFlow}\\
\partial_\Lambda X_\Lambda &= \frac{X_\Lambda^2}{\pi}\int_{p\in \text{FZ}}4\alpha^2\frac{Q_\Lambda^2\alpha^2 - \Lambda (\partial_\Lambda Q_\Lambda^2)\alpha^2 - \frac{\Lambda^2}{4}}{\left[4Q_\Lambda^2\alpha^2 + \Lambda^2\right]^2}, \label{XFlow}
\end{align}
and
\begin{align}
\partial_\Lambda Y_\Lambda =& \frac{Y_\Lambda^2}{\pi}\int_{p\in FZ}\Bigg\{\frac{\Lambda\alpha^2(\partial_\Lambda Q_\Lambda)}{Q_\Lambda\left[4Q_\Lambda^2\alpha^2 + \Lambda^2\right]}-\frac{4Q_\Lambda^2\alpha^4 + \Lambda^2\alpha^2}{\left[4Q_\Lambda^2\alpha^2 + \Lambda^2\right]^2}\nonumber\\
& +\frac{\alpha(\partial_\Lambda Q_\Lambda)}{2Q_\Lambda^2}\left[\arctan\left(\frac{\Lambda}{2Q_\Lambda\alpha}\right) - \frac{\pi}{2}\right] \Bigg\}\,, \label{YFlow}
\end{align}
where $\alpha \equiv |\cos p_x + \cos p_y|$. Note that Katanin corrections to the running of the four-fermion couplings have been included, cf.~App.~\ref{FlowForce}. 
Eqs.~\eqref{QFlow} and~\eqref{XFlow} can now be solved by standard numerical techniques\footnote{The flow equations can of course also be solved exactly by means of the procedure introduced in Ref.~\onlinecite{Salmhofer2004}. This yields, however, the mean-field gap equation and does not convey information about cutoff dependencies etc. as discussed in the following (see also App.~\ref{Gapex} for details)}. 
In Fig.~\ref{RGGap}, the results for the order parameter $Q_\Lambda$, converted into $Q(T)$, are shown for different initial values $Q_{\Lambda\rightarrow\infty}$ and compared to the mean-field result for $\varphi = 0$, cf.~Fig.~\ref{MFResPlot}.

For $Q_{\Lambda\rightarrow\infty} \rightarrow 0$, the results for $T_c$ as well as $Q(T=0)$ converge towards the mean-field result as expected. The reproduction of the critical temperature hereby serves as a sanity check, since the reduced system, Eq.~\eqref{T0SymmBeta}, already provided this feature. The correct magnitude of the order parameter at vanishing temperature ensures that the interpretation of $Q_\Lambda$ as order parameter is meaningful.

Interestingly, the simple linear correspondence between temperature and RG scale does not hold in the symmetry broken regime anymore. As apparent from Fig.~\ref{RGGap}, the approach of $Q_{\Lambda\rightarrow 0}$ follows a different law compared to $Q_{\m{mf}}(T\rightarrow 0)$. A similar result has previously been found by the authors of Ref.~\onlinecite{Gersch2005} with respect to a charge density wave ordered system. In order to understand this behavior, we again have to think of the correspondence between $\Lambda$ and $T$ as being due to a simple integral approximation of the Matsubara sum. This works in the symmetric phase but breaks down in the symmetry broken regime. The relation may be rendered nonlinear inside the ordered phase due to the presence of the non-trivially $\Lambda$-dependent coupling $Q_\Lambda$ in the fermionic propagator.

This finding is important for the interpretation of finite-temperature observables obtained from a zero temperature pf-FRG analysis. As long as symmetries are asymptotically preserved (i.e. for $Q_{\Lambda\rightarrow\infty}\rightarrow 0$), a simple correspondence can be found. As soon as spontaneous symmetry breaking occurs, however, a linear interpretation of $\Lambda$ as an effective temperature is \emph{not possible} anymore. Reliable values for finite temperature observables such as, e.g., the magnetic susceptibility, can only be recovered by employing true Matsubara frequencies and adapted cutoff functions $\vartheta_\Lambda$ such as Eq.~\eqref{EnssReg}. We come back to this point in more detail in Sec.~\ref{SSBFRGTf}.

There is another detail of the results presented in Fig.~\ref{RGGap} that deserves attention. Comparing free energies, the mean field calculation shows that the phase with $\varphi = 0$ is not the true ground state, but rather the one with $\varphi = \frac{\pi}{2}$. Technically, our RG analysis has therefore not brought us into a true spin liquid phase but rather into a BZA resonating valence bond phase. 
This is understood as follows:
Both BZA and staggered-flux spin liquid phases are characterized by spontaneous breaking of the local U$(1)$ symmetry of the fermions. They even share the same critical temperature. The additional breakdown of lattice-translational invariance and/or isotropy associated with the spin liquid cannot be represented in the stripped-down ansatz,  Eq.~\eqref{T0sbGk}. 
The possibility of translation symmetry breaking could be implemented by the inclusion of nontrivial momentum dependencies~\cite{SalmHon2001,RevModPhys.84.299} of $Q_\Lambda, J_\Lambda, I_\Lambda$ into our scheme \footnote{We also note that a \emph{bosonic} spin representation yields an even better description of the ground state as the free energy found for the zero temperature \emph{homogeneous} state is considerably lower than even the $\varphi = \frac{\pi}{2}$ fermionic flux phase~\cite{ArovasAuerbach88}. It is therefore likely, that even the latter does not represent the true ground state, but rather an even more complicated spatio-temporal ordering structure.}.

Besides the possibility to fine-grain the momentum structure of couplings in the present approach, once real space pf-FRG schemes are developed which are able to efficiently deal with explicitly broken local symmetries, this problem will automatically be overcome by the very nature of the position space ansatz and its inherent structural complexity. For now, we leave this to future work. The goal of the present approach is, instead, to efficiently probe, identify and characterize the mechanism driving the general behavior of the system with respect to its symmetries.

\subsection{Magnetic channel}
\label{FMC}

One of the main reasons to employ pf-FRG directly, instead of a bosonized flow is the unbiased nature of the former. 
In a fully momentum and frequency dependent pf-FRG scheme, the structure of the four-Fermi vertex can faithfully monitor the evolution of complicated and even competing ordering processes~\cite{HusGieSalm2012,UebelHon2012,GieSalm2012,Licht2017}. In contrast, a Hubbard-Stratonovich transformation at the UV scale essentially focuses on one particular channel. Spontaneous symmetry breaking in a different one may go completely unnoticed and a possible competition between channels may be lost, at least partially, unless other Hubbard-Stratonovich fields are additionally considered\cite{PhysRevB.83.155125,PhysRevB.93.125119}.

The present ansatz of introducing a small symmetry breaking initial condition combines the advantages of both approaches. Access to the symmetry broken regime is provided just as in the bosonized system. Of course, a specific choice for the structure of the symmetry breaking terms $\Gamma_\Lambda^{\m{esb}}$ has to be made. When the full frequency and momentum structure of the fermion vertex are maintained, peculiarities of competing orders are still visible there. Moreover, in case $\Gamma_\Lambda^{\m{esb}}$ represents the wrong channel altogether, it fails to regularize the divergence of the fermion vertex function at finite scales. 
In fact, a divergence in the vertex despite the presence of an explicitly symmetry breaking term in the ansatz is therefore a very strong hint towards ordering in a different channel. One can then further explore the situation by including more/different symmetry breaking terms until the decisive one is found. In this sense, the unbiased nature of the fermionic flow is largely kept, while access to symmetry broken phases is provided nonetheless.

To illustrate this point, we investigate another potential symmetry breaking channel of the model in Eq.~\eqref{HeisenHam}. If the system would develop any kind of long-range magnetic ordering, the global SU$(N)$ symmetry would be spontaneously broken. In order to check whether this happens, a corresponding initial condition is devised instead of Eq.~\eqref{GESB}, reading
\begin{equation}
\label{GESBMag}
\Gamma_\Lambda^{\m{esb}} = \int_\tau \sum_{i,\alpha\beta} M_\Lambda f^\dagger_{i\alpha}f^{\phantom\dagger}_{i\beta} \,.
\end{equation}
Since it is spatially homogeneous, the ansatz in Eq.~\eqref{GESBMag} respects local U$(1)$ symmetry but breaks global SU$(N)$ as it is not a spin singlet. A straightforward calculation analogous to the spin-liquid case yields the flow equations for $J_\Lambda$ and $M_\Lambda$, see App.~\ref{FlowForce} for details. For any finite $N$, they are rather complicated, in particular since new spin structures such as $f^\dagger_\alpha f_\alpha f^\dagger_\beta f_\gamma$ or $f^\dagger_\alpha f_\beta f^\dagger_\gamma f_\delta$ for the quartic interaction terms are generated. When considering $N\rightarrow\infty$, however, these new structures are suppressed and we find
\begin{equation}
\label{MagRedFlow}
\partial_\Lambda M_\Lambda = 0,\quad \partial_\Lambda J_\Lambda = -J_\Lambda^2\partial_\Lambda \int_\omega \frac{\vartheta_v^2}{(i\omega)^2}.
\end{equation}
Any corrections to these remaining flow equations are of order $1/N$. Due to the vanishing of its flow, cf.~Eq.~\eqref{MagRedFlow}, the magnetization $M_\Lambda$ does not grow beyond its initial value. Furthermore, since the flow equation of $J_\Lambda$ is identical to the symmetric result, Eq.~\eqref{T0SymmBeta}, the divergence at $\Lambda_{\m{sb}}$ still occurs, prohibiting access to the infrared regime. As expected, it can be concluded that the  ansatz in Eq.~\eqref{GESBMag} does not capture the correct symmetry breaking pattern.\bigskip

Summarizing, our minimalistic momentum-space approach is able to identify the correct ordering channel among a few physically motivated, reasonable choices. It does not yet allow to resolve the momentum- or position-space structure in this ordering channel.

\section{Symmetry Broken Regime II: \boldmath$T>0$}
\label{SSBFRGTf}

The results of Sec.~\ref{SSBFRGT0} are of limited use only, as long as finite temperature observables cannot be extracted reliably. There are two obstacles to be overcome: The first is to extend the analysis to a true finite temperature description for the fermionic fields. If this is achieved, the second one is how to deal with the fermion number constraint which has been ignored almost completely, so far.

\subsection{Observables at finite \textit{T}}
\label{TfObs}

Let us begin with setting up a finite temperature description. Since the number constraint is effectively removed from the mean-field calculation of Sec.~\ref{AAMF}, anyway, we will be able to compare results directly.
On a formal level, the only thing that needs to be done is implementing the finite-temperature cutoff, Eq.~\eqref{EnssReg}, and compactifying the imaginary time axis with circumference $\beta = \frac{1}{T}$. 
Ignoring the flow of $Y_\Lambda$ as being of minor importance for the observables to be considered, the flow equations are then given by
\begin{subequations}
\label{fTFlows}
\begin{align}
\partial_\Lambda Q_\Lambda &= -\frac{X_\Lambda}{2}\int_{p\in FZ}\frac{\pi T^2\,Q_\Lambda\alpha^2\,\tilde{\omega}_\Lambda^2\tilde{\vartheta}_\Lambda}{\left[4Q_\Lambda^2\alpha^2\tilde{\vartheta}_\Lambda^2  + \pi^2T^2\tilde{\omega}_\Lambda^2\right]^2},\label{fTQFlow}\\
\partial_\Lambda X_\Lambda &=  X_\Lambda^2 \int_{p\in FZ}\alpha^2\Bigg[\frac{12\pi^2T^2\tilde{\omega}_\Lambda^2\tilde{\vartheta}_\Lambda^3 Q_\Lambda^2\alpha^2 - \pi^4 T^4 \tilde{\omega}_\Lambda^4\tilde{\vartheta}_\Lambda}{\pi \left[4Q_\Lambda^2\alpha^2\tilde{\vartheta}_\Lambda^2  + \pi^2T^2\tilde{\omega}_\Lambda^2\right]^3}\nonumber\\
&- \sum_n\frac{4 Q_\Lambda\omega_n^2\vartheta_\Lambda^4\alpha^2 - 16\vartheta_\Lambda^6 Q_\Lambda^3 \alpha^4}{\left[4\vartheta_\Lambda^2Q_k^2\alpha^2 + \omega_n^2\right]^3}\partial_\Lambda Q_\Lambda\Bigg]\label{fTXFlow}.
\end{align}
\end{subequations}
For conciseness, we introduced $\tilde{\omega}_\Lambda = \left(2\left\lfloor \frac{\Lambda}{2\pi T}\right\rfloor + 1\right)$ and $\tilde{\vartheta}_\Lambda = \left(1+\left\lfloor\frac{\Lambda}{2\pi T}\right\rfloor - \frac{\Lambda}{2\pi T}\right)$. For the largest part, the Matsubara sum was reduced to two terms and could thus be performed analytically. The summands of the Katanin contribution which appear in the last summand in Eq.~\eqref{fTXFlow} come with a factor $\partial_\Lambda Q_\Lambda$ instead of $\partial_\Lambda \vartheta_\Lambda$. 
Therefore, they do not experience this reduction and the Matsubara sum can not be performed analytically. This statement also holds for a smoothed version of the cutoff in Eq.~\eqref{EnssReg} due to its singularity structure in the complex plane.
A numerical evaluation of Eqs.~\eqref{fTFlows} is therefore much more costly as compared to the $T=0$ system. Due to the overall simplicity, though, the absolute cost is still rather small. Moreover, in case momentum and frequency dependence of the vertex functions is to be included in future work, analytic integration cannot be done in the $T=0$ case either. 
So, there is no true increase in numerical cost due to the use of the finite-$T$ cutoff.

\begin{figure}[t]
\centering
\includegraphics[width = 90mm]{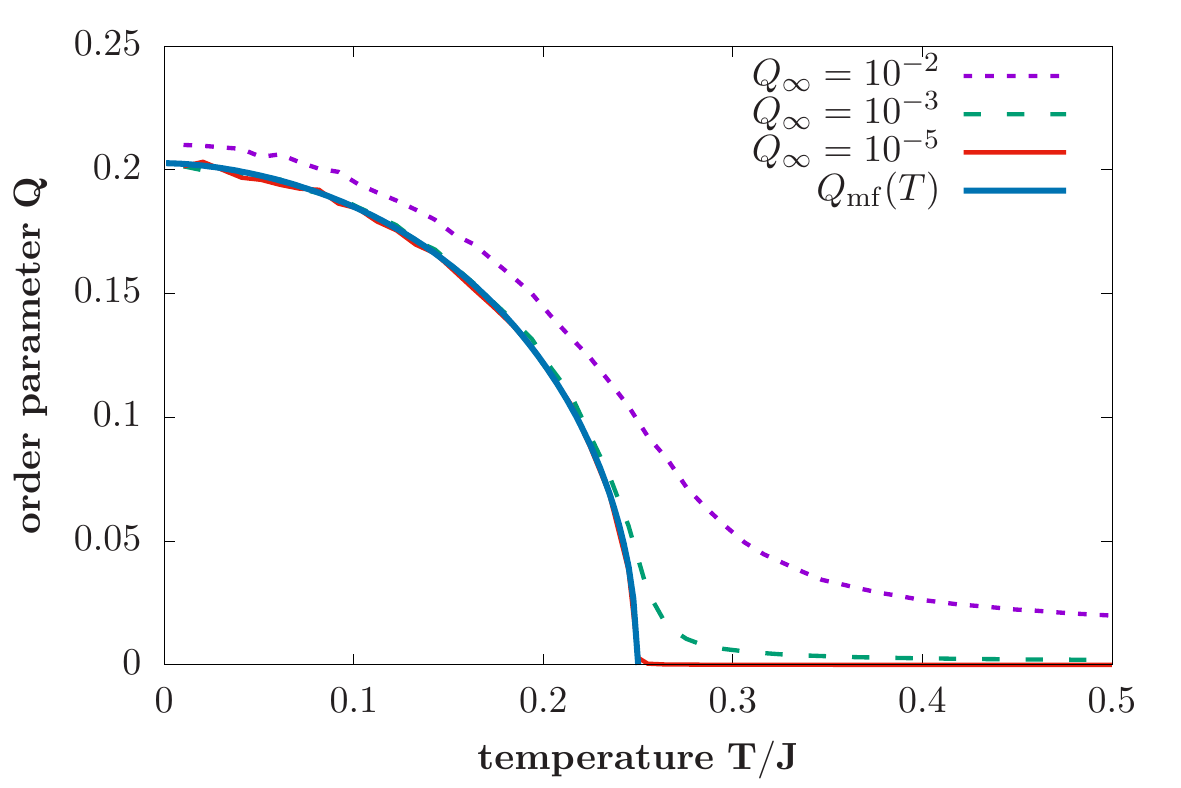}
\caption{Convergence of the {\bf finite-\textit{T} pf-FRG order parameter} magnitude from for different initial values $Q_{\Lambda\rightarrow\infty}$ to the mean field result $Q_{\m{mf}}$ for 100 Matsubara frequencies. For $Q_{\Lambda\rightarrow\infty} = 10^{-5}$, the pf-FRG results are practically indistinguishable from the direct mean field values. Slight wiggling is caused by numerical imprecision due to the truncated Matsubara sum.}
\label{QT}
\end{figure}

\begin{figure}[t]
\centering
\includegraphics[width = 90mm]{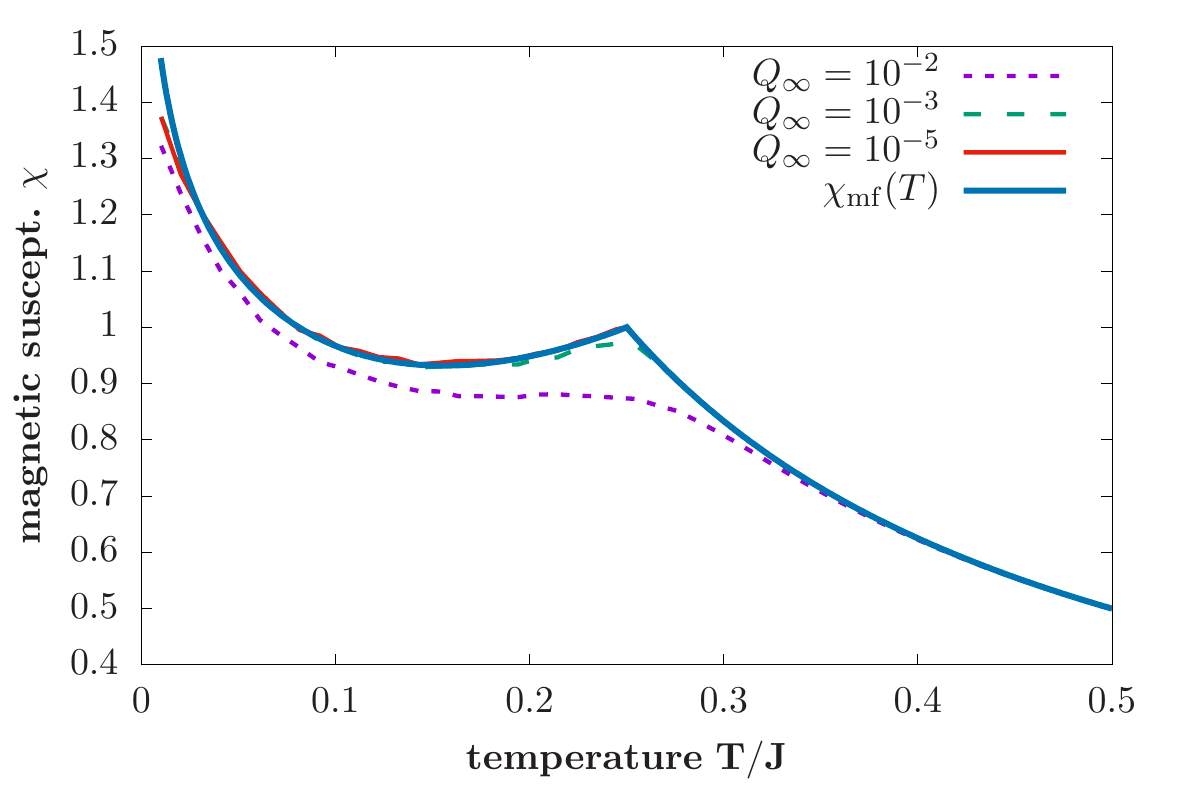}
\caption{Convergence of the {\bf finite-\textit{T} pf-FRG magnetic susceptibility} for different initial values $Q_{\Lambda\rightarrow\infty}$ to the mean field result $\chi_{\m{mf}}$ for 100 Matsubara frequencies.}
\label{chiT}
\end{figure}

As expected, the evaluation of Eqs.~\eqref{fTFlows} yields agreement with the mean-field results at percent level already for $\mathcal{O}(100)$ Matsubara frequencies, as exhibited in Figs.~\ref{QT} and~\ref{chiT}. This demonstrates the viability of the approach.

\subsection{Popov-Fedotov constraint}
\label{Popotov}

One of the main reasons to employ $T=0$ flows in previous pf-FRG approaches and to recover finite temperature results by the correspondence between $\Lambda$ and $T$ is the fact that the fermion number constraint is exactly fulfilled at vanishing temperature~\cite{PopovFedotov88}. 
When introducing actual finite fermion temperatures, unphysical states are populated which are not part of the original spin system but rather of the auxiliary fermion description, cf.~Eq.~\eqref{PFDef}. Thus, the results from Sec.~\ref{TfObs} are not necessarily solutions of the original spin model. 
There are different possibilities to implement the number constraint on the level of the path integral:

(1) A new field $\lambda$ is introduced and coupled to the fermion density. This generates a functional delta function enforcing the constraint, cf.~Sec.~\ref{AAMF}. However, the saddle-point solution $\lambda = 0$ puts this mechanism out of action. 
In the pf-FRG, Eq.~\eqref{lambdaconstraint} effectively represents a Yukawa interaction term between the constraint field $\lambda$ and the fermions. 
Higher-order contributions will be generated which would have to be taken into account
complicating a systematic analysis.
We therefore refrain from pursuing this approach.

(2) Alternatively, the constraint can be implemented by means of an imaginary chemical potential $\mu_{\m{ppv}} = i\frac{\pi T}{2}$ for SU$(2)$, according to Popov and Fedotov~\cite{PopovFedotov88}. It can be shown that the presence of such an object in the Hamiltonian cancels the unphysical states out of the partition function or any expectation values of (physical) observables~\cite{Binckle05}. Generalizations to different representations of SU$(N)$ and higher spin $S$ have been suggested in the literature~\cite{Kiselev2001}. 
This construction justifies the ignorance of the fermion number constraint in the $T=0$ case where $\mu_{\m{ppv}} = 0$. 
However, it also casts a shade of doubt on the finite temperature interpretation of flows stopped at some $\Lambda_{\m{stop}}$, as there is no way to implement the constraint for these.

An additional advantage of the finite-temperature regularization scheme, Eq.~\eqref{EnssReg}, is that the implementation of Popov-Fedotov-type constraints by means of an imaginary chemical potential becomes feasible.
For any finite $N$, the corresponding $\mu_{\m{ppv}}$ may be introduced, explicitly. 
To illustrate the simplicity of their treatment, let us consider flow equations for the case $N=2$. Provided a system of $\beta$~functions for two- and four-point functions, $\partial_\Lambda Q_\Lambda$ and $\partial_\Lambda X_\Lambda$, respectively, the modified RG equations are given by
\begin{subequations}
\label{PopotovFlow}
\begin{align}
\partial_\Lambda Q_\Lambda^{\m{ppv}} &= \frac{\partial_\Lambda Q_\Lambda}{2}\bigg|_{\tilde{\omega}_k = \tilde{\omega}_{k,1}} + \frac{\partial_\Lambda Q_\Lambda}{2}\bigg|_{\tilde{\omega}_k = \tilde{\omega}_{k,2}},\\
\partial_\Lambda X_\Lambda^{\m{ppv}} &= \frac{\partial_\Lambda X_\Lambda}{2}\bigg|_{\substack{\tilde{\omega}_k = \tilde{\omega}_{k,1}\\
 \omega_n = \omega_n^{\mu}}} 
 + \frac{\partial_\Lambda X_\Lambda}{2}\bigg|_{
 \substack{\tilde{\omega}_k = \tilde{\omega}_{k,2}\\
  \omega_n = \omega_n^{\mu}}}\,,
\end{align}
\end{subequations}
were $\tilde{\omega}_{k,1} = \left(2\left\lfloor \frac{k}{2\pi T}\right\rfloor + \frac{3}{2}\right)$, $\tilde{\omega}_{k,2} = \left(2\left\lfloor \frac{k}{2\pi T}\right\rfloor + \frac{1}{2}\right)$ and $\omega_n^\mu = \omega_n + \mu^{\m{ppv}}$.
At least for small and finite $N$ an exact treatment of the number constraint is thus easily achieved. The limit $N\rightarrow\infty$ is more involved as a continuous distribution of imaginary chemical potentials must be employed~\cite{Kiselev2001}. We therefore leave an application to the present system for future work.

\section{Conclusions}
\label{Conclusions}

To summarize, we have introduced a generalized pf-FRG approach that allows us to access the low-temperature physics of  SU$(N)$ Heisenberg models, both in the symmetric and symmetry broken regimes.
For the square lattice SU$(N)$ Heisenberg model we calculated the order parameter and magnetic susceptibility in the large-$N$ limit for arbitrary temperatures, which we carefully compared to established results. 
We found that the commonly employed linear correspondence between the RG parameter $\Lambda$ and temperature $T$ does not hold anymore in the symmetry broken regime.
Introducing a finite-$T$ regularization, we overcome the inability of the $T=0$ pf-FRG approach  to correctly predict finite temperature observables for any system that exhibits spontaneous symmetry breaking.
An additional decisive feature of our approach is the introduction of a symmetry-breaking term which by itself induces only a minimal bias, as opposed to the common strategy of a single-channel Hubbard-Stratonovich transformation in the fermion interaction. 
In order to illustrate this, we analyzed another symmetry-breaking initial condition, breaking the global SU$(N)$ instead of the local U$(1)$ symmetry. 
Indeed, we found that the fermionic divergence is not regularized in this case, confirming that an inappropriate choice of symmetry breaking initial condition does not bias the investigation.
We also devised a strategy to properly handle the fermion number constraint at finite temperatures, supplementing the action by a Popov-Fedotov-type~\cite{PopovFedotov88} imaginary chemical potential. 
Within this framework, fulfilling the constraint exactly is easy for $N=2$ and we leave the generalization of this procedure to arbitrary $N$ for future work.

The technical developments presented in this manuscript and its companion paper~[\onlinecite{RSP17}] set the stage for future  investigations of spin liquid physics within the pf-FRG approach. 
As a general strategy, we suggest a two-step scheme to investigate a model of interest:
(1)~Quick determination of the symmetry-breaking properties by the maximally stripped-down pf-FRG approach presented here.
While this gives a strong hint on the leading physical instability, this simplified approach may not resolve more subtle aspects  of the spin liquid phase (as we have seen for the spatio-temporal structure of the order parameter in the case study of the manuscript at hand).
 (2)~In-depth analysis of the identified channel by means of fine grained position or momentum space pf-FRG, as developed in our companion manuscript, to resolve the fine-print of particular spin liquid phases.

A future challenge will be to fully incorporate gauge degrees of freedom into the pf-FRG approach.
To do so, we expect to strongly benefit from the simplicity of the stripped-down, momentum-space approach developed in this manuscript.
This will open up the possibility to directly link the emergent gauge theory description of spin liquid ground states to their parent lattice spin models. 

\begin{acknowledgments}
We thank J. Reuther for discussions.
This work was partially supported by the DFG within the CRC 1238 (project C03). 
F.L.B. thanks the Bonn-Cologne Graduate School of Physics and Astronomy (BCGS) for support.  
S.D. and D.R. acknowledge support by the German Research Foundation (DFG) 
through the Institutional Strategy of the University of Cologne within the German Excellence Initiative (ZUK 81).
D.R. is supported, in part, by the NSERC of Canada.
\end{acknowledgments}

\appendix

\section{Flow equations for the symmetry-broken regime}
\label{FlowForce}

In the following, we provide details of the computation leading to the flow Eqs.~\eqref{QFlow}, \eqref{XFlow} and \eqref{YFlow}. After having derived those, we consider the limit $Q_{\Lambda\rightarrow\infty}\rightarrow 0$ to reproduce the symmetric flow Eq.~\eqref{T0SymmBeta}.
The starting point is the ansatz in Eq.~\eqref{T0sbGk} for the effective average action. In order to be able to explicitly evaluate the projection rules, Eqs.~\eqref{T0sbProj}, the r.h.s. of the Wetterich Eq.~\eqref{Wetterhofer} will be expanded as
\begin{equation}
\label{ExpandWetterich}
\partial_\Lambda \Gamma_\Lambda = C + \frac{1}{2}\tilde{\partial}_\Lambda\m{STr}\sum_{n=1}^\infty\frac{(-1)^{n+1}}{n}\left[\mathcal{P}_\Lambda^{-1}\mathcal{F}_\Lambda\right]^n.
\end{equation} 
Here, $C$ comprises field-independent and bilinear terms that will not be important for our considerations and $\tilde{\partial}_\Lambda$ is a scale derivative that acts on the $\Lambda$ dependence of the cutoff function, only. While $\mathcal{P}_\Lambda^{-1}$ is the usual regulated propagator, $\mathcal{F}_\Lambda$ is the field-dependent contribution to the second functional derivative of $\Gamma_\Lambda$. While the flow equation for $Q_\Lambda$ is extracted from the $n=1$ term of Eq.~\eqref{ExpandWetterich}, the flow of the quartic couplings $I_k$ and $J_k$ is encoded in the $n=2$ contribution.

Performing the functional derivative and applying the single-mode projection $\Psi^{(\dagger)}_{a\alpha} = \Psi^{(\dagger)}_\alpha \delta(\omega_a)\delta(p_a)$ yields
\begin{align}
\mathcal{P}_\Lambda^{-1} &= -\frac{\delta_{\gamma\delta}\delta(\omega-\nu)\delta(p-q)}{\omega^2 + 4\vartheta_\Lambda^2Q_\Lambda^2\tilde{\alpha}^2}\begin{pmatrix} 0 & -P_f^\ast\\
P_f & 0 \end{pmatrix},
\end{align}
where $P_f=i\omega \vartheta_\Lambda\mathbbm{1} + 2Q_\Lambda\tilde{\alpha}\vartheta_\Lambda^2\sigma_x$ and $\tilde{\alpha} = \cos(p_x) + \cos(p_y)$ and $\alpha = |\tilde{\alpha}|$.
Further
\begin{align}
\label{Fk}
\mathcal{F}_\Lambda = \frac{2\tilde{\alpha}}{N} \delta_{\gamma\delta}\delta(\omega-\nu)\delta(p-q) \begin{pmatrix} 0 & O_V \\ -O_V & 0 \end{pmatrix},
\end{align}
with $O_V=O_I+O_J$ and
\begin{equation}
\begin{aligned}
O_I &= I_\Lambda\left[\sigma^+(\Psi^\dagger_\alpha\sigma^+\Psi^{\phantom\dagger}_\alpha) + \sigma^-(\Psi^\dagger_\alpha\sigma^-\Psi^{\phantom\dagger}_\alpha)\right]\\
O_J &= J_\Lambda\left[\sigma^+(\Psi^\dagger_\alpha\sigma^-\Psi^{\phantom\dagger}_\alpha) + (\Psi^\dagger_\alpha\sigma^+\Psi^{\phantom\dagger}_\alpha)\sigma^-\right].
\end{aligned}
\end{equation}
For $\mathcal{F}_\Lambda$, only leading order terms in $N$ have been kept.
Now straightforward matrix multiplication yields
\begin{align}
\m{STr} \left[\mathcal{P}_\Lambda^{-1}\mathcal{F}_\Lambda\right] &= A_\Lambda \Psi^\dagger_\alpha\sigma_x\Psi^{\phantom\dagger}_\alpha\\
\m{STr} \left[\left(\mathcal{P}_\Lambda^{-1}\mathcal{F}_\Lambda\right)^2\right] &= B_\Lambda \left[(\Psi^\dagger_\alpha\sigma^+\Psi_\alpha)^2 + (\Psi^\dagger_\alpha\sigma^-\Psi_\alpha)^2 \right]\nonumber\\
&\quad+ C_\Lambda (\Psi^\dagger_\alpha\sigma^+\Psi_\alpha)(\Psi^\dagger_\alpha\sigma^-\Psi_\alpha).
\end{align}
Here, $A_\Lambda$, $B_\Lambda$ and $C_\Lambda$ are the coefficients to be used in the projections rules, Eqs.~\eqref{T0sbProj}, to extract the flow equations.
Applying the projection rules based on the expansion for $\Gamma_\Lambda$, Eq.~\eqref{ExpandWetterich} directly results in flow equations that do not include the Katanin improvement yet. In order to achieve this, it is necessary to replace $\tilde{\partial}_\Lambda \,\rightarrow\, \tilde{\partial}_\Lambda + (\partial_\Lambda Q_\Lambda)\partial_{Q_\Lambda}$ for the projection on $\partial_\Lambda I_\Lambda$ and $\partial_\Lambda J_\Lambda$.
Taking this modification into account the flow equations for the couplings are finally given by
\begin{widetext}
\begin{subequations}
\begin{align}
\partial_\Lambda Q_\Lambda & = -\frac{1}{2\pi}\int_{p\in \text{FZ}}\frac{Q_\Lambda\alpha^2(J_\Lambda+I_\Lambda)}{4Q_\Lambda^2\alpha^2+\Lambda^2}\,,\label{QRawEq}\\[10pt]
\partial_\Lambda I_\Lambda & = \frac{1}{\pi}\int_{p\in \text{FZ}}\frac{2Q_\Lambda^2\alpha^4(J_\Lambda^2+I_\Lambda^2) - \Lambda^2\alpha^2J_\Lambda I_\Lambda}{\left[4Q_\Lambda^2\alpha^2+\Lambda^2\right]^2}-\frac{\partial_\Lambda Q_\Lambda}{2\pi}\int_{p\in \text{FZ}}\alpha\Bigg\{4\Lambda Q_\Lambda\alpha^3\frac{J_\Lambda^2+I_\Lambda^2+2J_\Lambda I_\Lambda}{\left[4\alpha^2Q_\Lambda^2+\Lambda^2\right]^2} \label{IRawEq}\\
&\quad + \frac{\Lambda\alpha}{2Q_\Lambda}\frac{J_\Lambda^2+I_\Lambda^2 -2J_\Lambda I_\Lambda}{\left[4\alpha^2Q_\Lambda^2+\Lambda^2\right]} + \frac{J_\Lambda^2+I_\Lambda^2 - 2J_\Lambda I_\Lambda}{4Q_\Lambda^2}\left[\arctan\left(\frac{\Lambda}{2\alpha Q_\Lambda}\right)-\frac{\pi}{2}\right]\Bigg\}\,, \nonumber\\[10pt]
\partial_\Lambda J_\Lambda & = \frac{1}{2\pi}\int_{p\in \text{FZ}}\frac{8Q_\Lambda^2\alpha^4J_\Lambda I_\Lambda - \Lambda^2\alpha^2(J_\Lambda^2+I_\Lambda^2)}{\left[4Q_\Lambda^2\alpha^2+\Lambda^2\right]^2}-\frac{\partial_\Lambda Q_\Lambda}{4\pi}\int_{p\in \text{FZ}}\alpha\Bigg\{8\Lambda Q_\Lambda\alpha^3\frac{J_\Lambda^2+I_\Lambda^2+2J_\Lambda I_\Lambda}{\left[4\alpha^2Q_\Lambda^2+\Lambda^2\right]^2} \label{JRawEq}\\
&\quad- \frac{\Lambda\alpha}{Q_\Lambda}\frac{J_\Lambda^2+I_\Lambda^2-2J_\Lambda I_\Lambda}{\left[4\alpha^2Q_\Lambda^2+\Lambda^2\right]} - \frac{J_\Lambda^2+I_\Lambda^2 - 2J_\Lambda I_\Lambda}{2Q_\Lambda^2}\left[\arctan\left(\frac{\Lambda}{2\alpha Q_\Lambda}\right)-\frac{\pi}{2}\right]\Bigg\}\,.\nonumber
\end{align}
\end{subequations}
\end{widetext}
The peculiar structure of the occurrence of $I_\Lambda$ and $J_\Lambda$ is already indicative of the transformation $X_\Lambda = J_\Lambda + I_\Lambda$, $Y_\Lambda = I_\Lambda - J_\Lambda$. Indeed, performing this transformation, the decoupled flow Eqs.~\eqref{QFlow}, \eqref{XFlow} and \eqref{YFlow} are obtained.

Let us now consider the limit $Q_{\Lambda\rightarrow\infty}\rightarrow 0$. 
Since $\partial_\Lambda Q_\Lambda \sim Q_\Lambda$ this yields $Q_\Lambda = 0$ throughout the flow. 
As the initial value $I_{\Lambda\rightarrow \infty} = 0$, we have $\partial_\Lambda I_\Lambda = 0$, as well because the corresponding $\beta$ function, Eq.~\eqref{IRawEq}, contains only components $\sim Q_\Lambda$, $\sim \partial_\Lambda Q_\Lambda$ and $\sim I_\Lambda$. Consequently, only
\begin{equation}
\partial_\Lambda J_\Lambda = -\frac{1}{2\pi}\int_{p\in \text{FZ}} \alpha^2\frac{I_\Lambda^2}{\Lambda^2}
\end{equation}
persists, which is equivalent to the zero-temperature symmetric flow Eq.~\eqref{T0SymmBeta} as it should be.

\section{Gap equation from the RG flow}
\label{Gapex}

It can be proven quite generally that the mean field gap equation can be derived from the Katanin-improved flow equations~\cite{Salmhofer2004,RSP17}. Here, we will demonstrate this procedure for our system. The derivation is particularly simple since neither the fermionic couplings nor the self-energy contribution $Q_\Lambda$ retain any dependencies on momenta or frequencies. While also serving as a sanity check for our results, this relation nicely illustrates the influence of the regularization scheme.
Let us introduce a shorthand notation for the flow Eqs.~\eqref{QFlow} and~\eqref{XFlow} reading
\begin{align}
\partial_\Lambda Q_\Lambda \equiv X_\Lambda \tilde{\partial}_\Lambda g_\Lambda\,,\quad
\partial_\Lambda X_\Lambda \equiv -X_\Lambda^2 \partial_\Lambda f_\Lambda\label{sXFlow}.
\end{align}
For Eq.~\eqref{sXFlow}, there is an exact (albeit implicit) solution,
\begin{equation}
\label{XSoln}
X_\Lambda = X_\infty - X_\infty X_\Lambda f_\Lambda.
\end{equation}
Plugging Eq.~\eqref{XSoln} into the second identity in Eq.~\eqref{sXFlow},
\begin{equation}
\begin{aligned}
\partial_\Lambda Q_\Lambda &= X_\infty \tilde{\partial}_\Lambda g_\Lambda - X_\infty (X_\Lambda \tilde{\partial}_\Lambda g_\Lambda) f_\Lambda\\
&= X_\infty\tilde{\partial}_\Lambda g_\Lambda - X_\infty(\partial_\Lambda Q_\Lambda) f_\Lambda\,.
\end{aligned}
\end{equation}
Conveniently, it holds that $\partial_{Q_\Lambda}g_\Lambda=-f_\Lambda$
and therefore 
\begin{align}
\tilde{\partial}_\Lambda g_\Lambda - (\partial_\Lambda Q_\Lambda) f_\Lambda = \tilde{\partial}_\Lambda g_\Lambda + (\partial_\Lambda Q_\Lambda)\partial_{Q_\Lambda} g_\Lambda = \partial_\Lambda g_\Lambda.\nonumber
\end{align}
Finally, we obtain
\begin{equation}
\label{FormdkGap}
\partial_\Lambda Q_\Lambda = X_\infty \partial_\Lambda g_\Lambda.
\end{equation}
Since the left and right hand sides of Eq.~\eqref{FormdkGap} are total derivatives, the differential equation can be integrated directly,
\begin{equation}
\label{FormGap}
Q_0 - Q_\infty = X_\infty(g_0 - g_\infty) \,\overset{Q_\infty \rightarrow 0}{\Longrightarrow}\, Q_0 = X_\infty g_0.
\end{equation}
For the last step, it has been assumed that the explicitly symmetry breaking initial condition $Q_\Lambda$ has been sent to zero. Since $g_\Lambda\sim Q_\Lambda$, it is $g_\Lambda = 0$ as well. Notice furthermore that $\vartheta_0 = 1$ almost everywhere. 
Eq.~\eqref{FormGap} is the well-known mean-field gap equation.


\bibliography{largeN}

\end{document}